# VECTOR CONTINUED FRACTIONS USING A GENERALISED INVERSE

Roger Haydock and C.M.M. Nex*

Department of Physics

University of Oregon, Eugene, OR 97403-1274

USA

Geoffrey Wexler, Cavendish Laboratory

Madingley Road, Cambridge CB3 0HE

UK

**Abstract**: A real vector space combined with an inverse (involution) for vectors is sufficient to define a vector continued fraction whose parameters consist of vector shifts and changes of scale. The choice of sign for different components of the vector inverse permits construction of vector analogues of the Jacobi continued fraction. These vector Jacobi fractions are related to vector and scalar-valued polynomial functions of the vectors, which satisfy recurrence relations similar to those of orthogonal polynomials. The vector Jacobi fraction has strong convergence properties which are demonstrated analytically, and illustrated numerically.

* Permanent address: 31 Roft Street, Oswestry SY11 2EP, UK.

## 1. Continued Fractions

An ordinary continued fraction is defined as repeated division and addition of numbers:

$$A/B + C/D + E/F + \ldots, \qquad (1)$$

where the division is by everything to the right of the slash. Continued fractions have attracted interest because they can provide rapidly convergent approximations for various arithmetic quantities. For a survey of continued fractions see Ref. 1.

Inclusion of a variable in continued fractions produces functions of a complex variable, for example a Jacobi continued fraction with the complex variable z, when the $\{a_n\}$ are all real and the $\{\beta_n\}$ are all positive in,

$$1/z - a_0 - \beta_1/z - a_1 - \beta_2/z - a_2 - \beta_3/z - \ldots, \qquad (2)$$

This continued fraction is closely related to orthogonal polynomials [2] and Gaussian quadrature [3].

The broad motivation for this work is the approximation of distributions and functions on vector spaces, analogous to the approximation of weight distributions and smooth functions by Jacobi fractions and the polynomials associated with them. In this work, we develop a vector continued fraction which we show has convergence properties similar to those of the Jacobi fraction.

While vector division might seem necessary for a vector continued fraction, we find interesting properties using only a generalised inverse of a vector. It seems that a real vector space together with an inverse for vectors is the minimum algebraic structure necessary to define a continued fraction.

This paper begins with the definition of a generalised inverse for vectors and then applies this to the vector generalisation of continued fractions of the Jacobi type. The next three Secs. develop polynomial functions of vectors, the associated Christofel-Darboux identity, and a theory of the convergence of these fractions. The penultimate Sec. of the paper contains analytic and numerical examples of convergence and other properties of these vector continued fractions, while the final Sec. contains remarks about quadrature and geometric algebra.

## 2. Definition of an Inverse for a Vector

The inverse (involution) $1/\mathbf{Z}$ of a vector $\mathbf{Z}$ has the property that the inverse of the inverse, $1/(1/\mathbf{Z})$ is the original vector $\mathbf{Z}$. A simple multiplicative inverse of a vector $\mathbf{Z}$ can be defined in terms of the magnitude $|\mathbf{Z}|$ of $\mathbf{Z}$ and a self-inverse (symmetric orthogonal) transformation $\sigma$,

$$1/\mathbf{Z} = \sigma \mathbf{Z} / |\mathbf{Z}|^2. \tag{3}$$

(We define this inverse in terms of a norm although that is not necessary in general.) Because $\sigma$ is self-inverse, it has a complete set of eigenvectors, and each eigenvalue is $\pm 1$. For $\mathbf{Z}$ one-dimensional, the inverse in Eq. 3 reduces to the usual inverse for real numbers if $\sigma$ is $+1$, and to the usual inverse for imaginary numbers if $\sigma$ is $-1$. For multi-dimensional $\mathbf{Z}$, components of eigenvectors with eigenvalue $+1$ (the positive eigenspace), behave like real numbers, and those of eigenvectors with eigenvalue $-1$ (the negative eigenspace), behave like imaginary numbers. For $\mathbf{Z}$ two-dimensional with $\sigma$ having eigenvalues $+1$ and $-1$, this inverse is the same as the reciprocal of a complex number, independent of complex multiplication; and for $\mathbf{Z}$ four-dimensional with $\sigma$ having eigenvalues $+1, -1, -1, -1$, this inverse is the same as the reciprocal of a quaternion, independent of quaternion multiplication.

This inverse is related to the Moore-Penrose inverse [4]. It was the basis of a mechanical linkage called the Peaucellier inversor [5], and is a special case of the Cremona [6] transformation. As with complex numbers, the zero-vector $\mathbf{0}$ can be given a unique inverse, the infinite vector $\infty$.

### 3. A Vector Continued Fraction of Jacobi Type

Using the vector inverse defined above, continued fractions of the Jacobi type can be generalised for vectors of dimension greater than two, provided that $\sigma$ has at least one positive eigenvector and at least one negative eigenvector. In order to demonstrate the convergence properties of these fractions, we choose one negative eigenvector of $\sigma$ as the embedding eigenvector, denoted in what follows by **y** and analogous to the role of the root of -1 for ordinary Jacobi fractions. It is also convenient to subsume $\sigma$ in the definition of the conjugate **Z**\* of the vector **Z**,

$$\mathbf{Z}^* = \sigma \, \mathbf{Z}, \tag{4}$$

which reduces to the complex conjugate when $\sigma$ has one positive and one negative eigenvalue.

For shifts $\{\mathbf{A}_n\}$ which have no components of **y**, scales $\{\beta_n\}$ which are all positive real numbers, and a variable vector **Z**, the Jacobi form of the vector continued fraction (JVF) is,

$$\mathbf{R}(\mathbf{Z}) = \beta_0 \,/\mathbf{Z} - \mathbf{A}_0 - \beta_1 \,/\mathbf{Z} - \mathbf{A}_1 - \beta_2 \,/\mathbf{Z} - \mathbf{A}_2 - ... - \beta_n \,/\mathbf{Z} - \mathbf{A}_n - ..., \tag{5}$$

where the slash indicates the inverse of the entire expression to its right. It is shown below that this has properties similar to the Jacobi fraction. Note that the embedding eigenvector **y** may be chosen to be any vector in the negative eigenspace of $\sigma$ provided that it has no components of the shifts $\{\mathbf{A}_n\}$. This does not preclude the shifts $\{\mathbf{A}_n\}$ from having components in the negative eigenspace of $\sigma$, it simply precludes **y** from being one of those eigenvectors, and requires that there be at least one vector in the negative eigenspace of $\sigma$ for which the shifts $\{\mathbf{A}_n\}$ have no component.

The JVF has an orientation which is the analogue of the Herglotz [7]

property of the Jacobi fraction, that the **y**-component of the JVF is opposite in sign to that of **Z**. First separate the vector space containing **Z** into two half spaces, $\Gamma^+$ with a positive component of **y** and $\Gamma^-$ with a negative component of **y**. We say that vectors with zero component of **y** lie on the boundary between the half spaces. Now consider a single level of the fraction, $\beta/\mathbf{Z}$-**A**-**T**, with **T** replacing the tail of the fraction. Assume that **Z** is in $\Gamma^+$, and suppose that **T**, which is itself a JVF, is in $\Gamma^-$, then **Z**-**A**-**T** is also in $\Gamma^+$, provided that **A** is on the boundary. Since **Z**-**A**-**T** is in $\Gamma^+$, its inverse and its inverse multiplied by a positive $\beta$, $\beta/\mathbf{Z}$-**A**-**T**, are both in $\Gamma^-$ which was what was to be shown. A similar argument shows that when **Z** is in $\Gamma^-$, the JVF is in $\Gamma^+$.

An important consequence of the above orientation property is that zeros and infinities of the JVF can occur only when **Z** is in the boundary between $\Gamma^+$ and $\Gamma^-$, making the behavior of the JVF on the boundary of particular interest. The argument is simple and analogous to that for ordinary Jacobi fractions. If **Z** is in $\Gamma^+$, then **R**(**Z**) must be in $\Gamma^-$, excluding **0** and $\infty$ from the possible values of **R**(**Z**) because neither **R**(**Z**) nor 1/**R**(**Z**) can take any value with a zero component of **y**. Since **y** can be any vector in the negative eigenspace containing no component of the shifts, this orientation property restricts the zeros and infinities of the JVF to values of **Z** which contain components of the shifts or are in the positive eigenspace.

The shift, inversion, and scaling at each level of a JVF, $\beta/\mathbf{Z}$-**A**-**T**, may be viewed as a transformation which takes the tail of the fraction, vector **T**, to another vector. Since $\sigma$ in Eq. 3 is orthogonal, the inverse defined from it preserves angles between vectors and so is conformal [8]. The shift part of the transformation **Z**-**A**-**T** clearly preserves angles, and multiplication by $\beta$ does not change angles, so each level of the JVF preserves angles making their composition, the whole fraction, a conformal transformation of its tail.

## 4. Polynomial Functions of Vectors

Finite Jacobi fractions can be expressed as ratios of orthogonal polynomials, and this carries over to JVFs to the extent that we show in this Sec. how the squared magnitude of a JVF can be expressed as the ratio of a numerator polynomial, related to orthogonal polynomials of the second kind [9], and a denominator polynomial, related to orthogonal polynomials of the first kind [9]. Polynomial functions of vectors are introduced here as a convenient way to specify the convergence properties of the JVFs, just as orthogonal polynomials are used for the convergence of Jacobi fractions.

In order to define polynomials, we need to define fragments of continued fractions. Given a set of shifts $\{\mathbf{A_n}\}$ and positive real scales $\{\beta_n\}$, define the forward fragments of the JVF as,

$$\mathbf{R}^{(m,n)}(\mathbf{Z}) = 1 / \mathbf{Z} - \mathbf{A}_m - \beta_{m+1} / \mathbf{Z} - \mathbf{A}_{m+1} - ... - \beta_n / \mathbf{Z} - \mathbf{A}_n, \qquad (6)$$

for $m \leq n$ and $\mathbf{0}$ otherwise, together with the reverse fragments,

$$\mathbf{S}^{(m,n)}(\mathbf{Z}) = 1 / \mathbf{Z} - \mathbf{A}_m - \beta_m / \mathbf{Z} - \mathbf{A}_{m-1} - ... - \beta_{n+1} / \mathbf{Z} - \mathbf{A}_n, \qquad (7)$$

for $m \geq n$ and $\mathbf{0}$ otherwise.

Different fragments have different zeros and infinities, but because $\mathbf{R}^{(m,n)}(\mathbf{Z})$ is $1/\mathbf{Z}-\mathbf{A}_m-\beta_{m+1}\mathbf{R}^{(m+1,n)}(\mathbf{Z})$, the zeros of $\mathbf{R}^{(m,n)}(\mathbf{Z})$ occur at the infinities of $\mathbf{R}^{(m+1,n)}(\mathbf{Z})$, and similarly for reverse fragments. Using this property, we construct a first kind (denominator) of scalar-valued, vector polynomial,

$$p_N(\mathbf{Z}) = \{1 / |\mathbf{R}^{(0,N-1)}(\mathbf{Z})|^2\} \{1 / |\mathbf{R}^{(1,N-1)}(\mathbf{Z})|^2\} \; ... \; \{1 / |\mathbf{R}^{(N-1,N-1)}(\mathbf{Z})|^2\}, \qquad (8)$$

which is a polynomial in the sense that it has no infinities for finite $\mathbf{Z}$. It is monic and of degree 2N in the sense that it behaves like $|\mathbf{Z}|^{2N}$ for large $|\mathbf{Z}|$, and its zeros are the infinities of $\mathbf{R}^{(0,N-1)}(\mathbf{Z})$.

Similarly, we may define a second kind (numerator) of scalar-valued vector polynomial,

$$q_{N-1}(\mathbf{Z}) = \{1/|\mathbf{R}^{(1,N-1)}(\mathbf{Z})|^2\}\{1/|\mathbf{R}^{(2,N-1)}(\mathbf{Z})|^2\} \ldots \{1/|\mathbf{R}^{(N-1,N-1)}(\mathbf{Z})|^2\}. \tag{9}$$

Again it has no infinities for finite $\mathbf{Z}$, is monic, and of degree $2(N-1)$ in the above sense. It has zeros at the infinities of $\mathbf{R}^{(1,N-1)}(\mathbf{Z})$ which are also the zeros of $\mathbf{R}^{(0,N-1)}(\mathbf{Z})$. Now dividing Eq. 9 by Eq. 8 gives,

$$q_{N-1}(\mathbf{Z})/p_N(\mathbf{Z}) = |\mathbf{R}^{(0,N-1)}(\mathbf{Z})|^2. \tag{10}$$

As with orthogonal polynomials, it is convenient to construct these vector polynomials from a recurrence whose parameters are those of the JVF in order of increasing index. In order to accomplish this we need to transform forward fragments into reverse fragments. This process begins with the identity,

$$|(\{1/[|\mathbf{R}^{(n+1,N-1)}(\mathbf{Z})|\ \mathbf{S}^{(n,0)}(\mathbf{Z})]\} - \beta_{n+1}\mathbf{R}^{(n+1,N-1)}(\mathbf{Z})/|\mathbf{R}^{(n+1,N-1)}(\mathbf{Z})|)|^2$$

$$= |(\{1/[|\mathbf{S}^{(n,0)}(\mathbf{Z})|\ \mathbf{R}^{(n+1,N-1)}(\mathbf{Z})]\} - \beta_{n+1}\mathbf{S}^{(n,0)}(\mathbf{Z})/|\mathbf{S}^{(n,0)}(\mathbf{Z})|)|^2, \tag{11}$$

which is simply that the magnitude of the sum of two vectors is unchanged by interchanging the orientations of the two vectors, given by $\mathbf{R}^{(n+1,N-1)}(\mathbf{Z})$ and $\mathbf{S}^{(n,0)}(\mathbf{Z})$. Factoring $1/|\mathbf{R}^{(n+1,N-1)}(\mathbf{Z})|^2$ from the LHS, $1/|\mathbf{S}^{(n,0)}(\mathbf{Z})|^2$ from the RHS of Eq. 11; replacing $1/\mathbf{R}^{(n+1,N-1)}(\mathbf{Z})$ by $\mathbf{Z}-\mathbf{A}_{n+1}-\beta_{n+2}\mathbf{R}^{(n+2,N-1)}(\mathbf{Z})$, and then replacing $\mathbf{Z}-\mathbf{A}_{n+1}-\beta_{n+1}\mathbf{S}^{(n,0)}(\mathbf{Z})$ by $1/\mathbf{S}^{(n+1,0)}(\mathbf{Z})$ gives,

$$|\{[1/\mathbf{S}^{(n,0)}(\mathbf{Z})] - \beta_{n+1}\mathbf{R}^{(n+1,N-1)}(\mathbf{Z})\}|^2 \{1/|\mathbf{R}^{(n+1,N-1)}(\mathbf{Z})|^2\}$$

$$= \{1/|\mathbf{S}^{(n,0)}(\mathbf{Z})|^2\}\,|\{[1/\mathbf{S}^{(n+1,0)}(\mathbf{Z})] - \beta_{n+2}\mathbf{R}^{(n+2,N-1)}(\mathbf{Z})\}|^2. \tag{12}$$

Starting from the left in Eq. 8, replace $1/|\mathbf{R}^{(0,N-1)}(\mathbf{Z})|^2$ by $|\{[1/\mathbf{S}^{(0,0)}(\mathbf{Z})]-\beta_1\mathbf{R}^{(1,N-1)}(\mathbf{Z})\}|^2$ using Eqs. 6 and 7, apply Eq. 12, and continue with each succeeding factor to obtain,

$$p_N(\mathbf{Z}) = \{1/|\mathbf{S}^{(0,0)}(\mathbf{Z})|^2\}\{1/|\mathbf{S}^{(1,0)}(\mathbf{Z})|^2\}\ldots\{1/|\mathbf{S}^{(N-1,0)}(\mathbf{Z})|^2\}. \tag{13}$$

Now the recurrence for the $\{p_n(\mathbf{Z})\}$ follows from the observation that

$$p_{N+1}(\mathbf{Z})\,p_N(\mathbf{Z}) = |p_N(\mathbf{Z})/\mathbf{S}^{(N,0)}(\mathbf{Z})|^2$$

$$= |\mathbf{Z}\,p_N(\mathbf{Z}) - A_N\,p_N(\mathbf{Z}) - \beta_N\,p_N(\mathbf{Z})\,\mathbf{S}^{(N-1,0)}(\mathbf{Z})|^2. \tag{14}$$

It is convenient to define a sequence of vector-valued functions which are shown below to be polynomials,

$$\mathbf{P}_{N+1}(\mathbf{Z}) = p_{N+1}(\mathbf{Z})\,\mathbf{S}^{(N,0)}(\mathbf{Z})^*. \tag{15}$$

The recurrence for these functions comes from replacing $p_{N+1}(\mathbf{Z})$ by $p_N(\mathbf{Z})/|\mathbf{S}^{(N,0)}(\mathbf{Z})|^2$ and noting that $\mathbf{S}^{(N,0)}(\mathbf{Z})^*/|\mathbf{S}^{(N,0)}(\mathbf{Z})|^2$ is $1/\mathbf{S}^{(N,0)}(\mathbf{Z})$, which is $(\mathbf{Z}-A_N)p_N(\mathbf{Z})-\beta_N\mathbf{P}_N(\mathbf{Z})^*$ so that,

$$\mathbf{P}_{N+1}(\mathbf{Z}) = (\mathbf{Z} - A_N)\,p_N(\mathbf{Z}) - \beta_N\,\mathbf{P}_N(\mathbf{Z})^*, \tag{16}$$

making it a vector polynomial for suitable initial conditions of the recurrence. The scalar-valued and vector-valued vector polynomials are then related by Eq. 14, which becomes,

$$p_{N+1}(\mathbf{Z}) \, p_N(\mathbf{Z}) = |\mathbf{P}_{N+1}(\mathbf{Z})|^2. \tag{17}$$

A similar derivation of the recurrence relations for the numerator polynomials $q_{N-1}(\mathbf{Z})$ can be avoided by observing that the numerator polynomials for $\mathbf{R}^{(0,N-1)}(\mathbf{Z})$ are also the denominator polynomials for $\mathbf{R}^{(1,N-1)}(\mathbf{Z})$, and so satisfy the same recurrence relations as the denominator polynomials, but with a shift in the index of the coefficients. Both sets of polynomials satisfy the initial conditions that $p_{-1}(\mathbf{Z})=q_{-1}(\mathbf{Z})=0$, $\mathbf{P}_0(\mathbf{Z})=\mathbf{Q}_0(\mathbf{Z})=\mathbf{0}$, $p_0(\mathbf{Z})=q_0(\mathbf{Z})=1$, together with the recurrence, Eq. 16, where $\mathbf{A}_{N+1}$ and $\beta_{N+1}$ replace $\mathbf{A}_N$ and $\beta_N$ in the case of the $\{q_N(\mathbf{Z})\}$. Equation 17 relates the scalar and vector-valued polynomials of both kinds. The relation between these vector polynomials and ordinary orthogonal polynomials is that the scalar-valued polynomial $p_N(\mathbf{Z})$ corresponds to the square of the ordinary denominator polynomial of degree N, and the vector-valued polynomial $\mathbf{P}_N(\mathbf{Z})$ corresponds to the product of denominator polynomials of degree N and N-1.

## 5. A Christoffel-Darboux Identity

The above definitions of vector polynomials lead to a relation between products of polynomials, similar to the Christoffel-Darboux identity [10], and used below to describe the convergence of JVFs. The identity is,

$$p_n(\mathbf{Z}) \, q_{n-1}(\mathbf{Z}) \, |[\mathbf{P}_{n+1}(\mathbf{Z})/p_n(\mathbf{Z})] - \mathbf{Q}_n(\mathbf{Z})/q_{n-1}(\mathbf{Z})|^2$$

$$= (\beta_n)^2 \, p_{n-1}(\mathbf{Z}) \, q_{n-2}(\mathbf{Z}) \, |[\mathbf{P}_n(\mathbf{Z})/p_{n-1}(\mathbf{Z})] - \mathbf{Q}_{n-1}(\mathbf{Z})/q_{n-2}(\mathbf{Z})|^2. \tag{18}$$

The first step in deriving this relation is to substitute the recurrence in Eq. 16 for $\mathbf{P}_{n+1}(\mathbf{Z})$ and $\mathbf{Q}_n(\mathbf{Z})$ to give,

$$p_n(\mathbf{Z}) \, q_{n-1}(\mathbf{Z}) \, |[\mathbf{P}_{n+1}(\mathbf{Z})/p_n(\mathbf{Z})] - \mathbf{Q}_n(\mathbf{Z})/q_{n-1}(\mathbf{Z})|^2$$

$$= (\beta_n)^2 \, p_n(\mathbf{Z}) \, q_{n-1}(\mathbf{Z}) \, |[\mathbf{P}_n(\mathbf{Z})^*/p_n(\mathbf{Z})] - \mathbf{Q}_{n-1}(\mathbf{Z})^*/q_{n-1}(\mathbf{Z})|^2. \qquad (19)$$

To get Eq. 18, normalise $\mathbf{P}_n(\mathbf{Z})$ and $\mathbf{Q}_{n-1}(\mathbf{Z})$ using Eq. 17, and then interchange the resulting unit vectors on the RHS of Eq. 18, which leaves the squared magnitude unchanged, and simplify.

## 6. Convergence of Jacobi Vector Fractions

One of the advantages of the Jacobi form for continued fractions is its simple convergence properties. As shown in Sec. 3, the Herglotz property of the JVF allows zeros and infinities only for $\mathbf{Z}$ with zero components of the embedding eigenvector $\mathbf{y}$. It is shown below that for $\mathbf{Z}$ which have non-zero components of $\mathbf{y}$, the value of the fraction is limited to a finite hypersphere, and that for infinite JVFs, the values of successive approximants lie within a sequence of nested, kissing hyperspheres.

An infinite JVF converges for a particular value of $\mathbf{Z}$ provided the above sequence of nested, kissing hyperspheres converges to a single point. This question is addressed here by calculating the dependence of the value of the fraction on the level at which the infinite tail of the fraction is approximated, extending the approach used for scalar continued fractions [11]. Such an approximation to the infinite JVF is the fraction,

$$\mathbf{R}_N(\mathbf{Z}, \mathbf{T}) = 1/\mathbf{Z} - \mathbf{A}_0 - \beta_1/\mathbf{Z} - \mathbf{A}_1 - \ldots - \beta_{N-1}/\mathbf{Z} - \mathbf{A}_{N-1} - \mathbf{T}, \qquad (20)$$

where the subscript N indicates the level at which the infinite tail of the fraction is replaced by some $\mathbf{T}$ which takes arbitrary values in one half space.

From the orientation property, discussed in Sec. 3, the sign of the component of **y** can be used to separate the vector space into two half spaces, $\Gamma^+$ for the positive components, and $\Gamma^-$ for the negative components, with the boundary between them having a zero component of **y**. If **Z** lies in $\Gamma^-$, then the value of the infinite JVF must lie in the opposite half-space $\Gamma^+$, and similarly, the part of the fraction replaced by **T** must also lie in $\Gamma^+$, which is the only restriction on the value of **T**. The same is true if the two half-spaces are interchanged. Given this, we can ask what is the image of $\Gamma^+$ under the mapping $1/\mathbf{Z}\text{-}\mathbf{A}_{N-1}\text{-}\mathbf{T}$? Since this mapping is an inversion through the point $\mathbf{Z}\text{-}\mathbf{A}_{N-1}$, it maps hyperspheres to hyperspheres. The half-space $\Gamma^+$ of allowed values of **T** is a hypersphere of infinite radius, so it is mapped to the interior of a hypersphere in $\Gamma^+$.

We can ask how an additional level of the fraction changes this? Now replace **T** by $\beta_N/\mathbf{Z}\text{-}\mathbf{A}_N\text{-}\mathbf{T}'$ where **T**' is the tail of a fraction with one extra explicit level. The image of **T**' due to $\beta_N/\mathbf{Z}\text{-}\mathbf{A}_N\text{-}\mathbf{T}'$ is, by the above argument, a hypersphere within $\Gamma^+$, and in this case the infinite vector in **T**' is mapped to **0** which lies on the boundary between half-spaces. Since the next level of the fraction $1/\mathbf{Z}\text{-}\mathbf{A}_{N-1}\text{-}\mathbf{T}$ maps $\Gamma^+$ to the interior of a hypersphere, it maps the image of **T**' to a hypersphere which lies inside the image of **T**. Since **0** is on the boundary of both the half-space and the image of $\beta_N/\mathbf{Z}\text{-}\mathbf{A}_N\text{-}\mathbf{T}'$, $1/\mathbf{Z}\text{-}\mathbf{A}_{N-1}$ lies both on the hypersphere which is the image of **T** and on the one which the image of **T**'. This is where the hyperspheres which are successive images of **T** and **T**' kiss. Starting this construction of hyperspheres with the first level of an infinite
continued fraction, Fig. 1 shows how each successive level limits the possible values of the fraction to a new hypersphere which lies inside the previous one and touches it at exactly one point. If the limiting hypersphere is a single point, the JVF converges for that **Z**.

We have now succeeded in dividing **T**-space, and its image $\mathbf{R}_N(\mathbf{Z}, \mathbf{T})$ under Eq. 20, into allowed and forbidden regions. In the case of **T**-space, the allowed

-11-

region is $\Gamma^+$ and the forbidden region is $\Gamma^-$; while the allowed region of $\mathbf{R}_N(\mathbf{Z}, \mathbf{T})$ is the interior of the hypersphere shown in Fig. 2, which is the image of $\Gamma^+$. To determine the radius of the allowed $\mathbf{R}_N(\mathbf{Z}, \mathbf{T})$-hypersphere, we calculate the difference between $|\mathbf{R}_N(\mathbf{Z}, \tau_{max})|$ and $|\mathbf{R}_N(\mathbf{Z}, \tau_{min})|$ which are respectively the distances from the origin to the furthest and nearest points on the hypersphere of allowed values of $\mathbf{R}_N(\mathbf{Z}, \mathbf{T})$, as shown in Fig. 2.

Consider hyperspheres in $\mathbf{R}_N(\mathbf{Z}, \mathbf{T})$-space centred at $\mathbf{R}_N(\mathbf{Z}, \mathbf{T})=\mathbf{0}$ and with constant $|\mathbf{R}_N(\mathbf{Z}, \mathbf{T})|$, which are images of hyperspheres in $\mathbf{T}$-space according to the properties of the JVF described above. For a given value of $|\mathbf{R}_N(\mathbf{Z}, \mathbf{T})|$, the corresponding hypersphere in $\mathbf{T}$-space satisfies,

$$|\mathbf{R}_N(\mathbf{Z}, \mathbf{T})|^2 = [q_{N-2}(\mathbf{Z})/p_{N-1}(\mathbf{Z})]$$

$$\times |\mathbf{Q}_{N-1}(\mathbf{Z})/q_{N-2}(\mathbf{Z}) - \mathbf{T}|^2 / |\mathbf{P}_N(\mathbf{Z})/p_{N-1}(\mathbf{Z}) - \mathbf{T}|^2, \qquad (21)$$

where the tail $\mathbf{T}$ is added to the shift $\mathbf{A}_{N-1}$ in the recurrences for $q_{N-1}(\mathbf{Z})$ and $p_N(\mathbf{Z})$, and the result substituted in Eq. 10. Equation 21 is a quadratic in $\mathbf{T}$ from which the centre $\tau_c$ and radius $t_T$ of the hypersphere in $\mathbf{T}$-space, whose image is the hypersphere of constant $|\mathbf{R}_N(\mathbf{Z}, \mathbf{T})|$, can be extracted by completing the square:

$$\tau_c = (\mathbf{Q}_{N-1}(\mathbf{Z}) - |\mathbf{R}_N(\mathbf{Z}, \mathbf{T})|^2 \mathbf{P}_N(\mathbf{Z}))/(q_{N-2}(\mathbf{Z}) - p_{N-1}(\mathbf{Z})|\mathbf{R}_N(\mathbf{Z}, \mathbf{T})|^2), \qquad (22)$$

and,

$$t_T^2 = |\tau_c|^2 + \{(q_{N-2}(\mathbf{Z})|\mathbf{R}_N(\mathbf{Z}, \mathbf{T})|^2|\mathbf{P}_N(\mathbf{Z})|^2 - p_{N-1}(\mathbf{Z})|\mathbf{Q}_{N-1}(\mathbf{Z})|^2)$$

$$/(p_{N-1}(\mathbf{Z}) q_{N-2}(\mathbf{Z})^2 - p_{N-1}(\mathbf{Z})^2 q_{N-2}(\mathbf{Z})|\mathbf{R}_N(\mathbf{Z}, \mathbf{T})|^2)\}. \qquad (23)$$

The condition on $\tau_c$ and $t_T$ that the image hypersphere have radius either

$|\mathbf{R}_N(\mathbf{Z}, \tau_{max})|$ or $|\mathbf{R}_N(\mathbf{Z}, \tau_{min})|$ is that the hypersphere in **T**-space should touch $\Gamma^+$, which is that,

$$t_T = \tau_c \cdot \mathbf{y}. \tag{24}$$

Equation 24 combined with Eqs. 22 and 23 gives a single quadratic equation for $|\mathbf{R}_N(\mathbf{Z}, \tau_{max/min})|$ and the difference between its two roots is twice the radius $\rho_N(\mathbf{Z})$ of the hypersphere of allowed values for $\mathbf{R}_N(\mathbf{Z}, \mathbf{T})$,
$2\rho_N(\mathbf{Z}) = |\mathbf{R}_N(\mathbf{Z}, \tau_{max})| - |\mathbf{R}_N(\mathbf{Z}, \tau_{min})|$. The square of the difference between the roots can be calculated directly from the coefficients of the quadratic to give,

$$\rho_N(\mathbf{Z})^2 = p_{N-1}(\mathbf{Z}) \, q_{N-2}(\mathbf{Z}) \, |\mathbf{P}_N(\mathbf{Z})/p_{N-1}(\mathbf{Z}) - \mathbf{Q}_{N-1}(\mathbf{Z})/q_{N-2}(\mathbf{Z})|^2$$

$$/[2 \, \mathbf{y} \cdot \mathbf{P}_N(\mathbf{Z})\}]^2, \tag{25}$$

in terms of the vector polynomials.

The above expression can be simplified using the identity in Eq. 18 repeatedly to get,

$$\rho_N(\mathbf{Z})^2 = (\beta_{N-1})^2 \, (\beta_{N-2})^2 \, ... \, (\beta_2)^2 \, (\beta_1)^2 \, /[2 \, \mathbf{y} \cdot \mathbf{P}_N(\mathbf{Z})]^2. \tag{26}$$

A related, though less elegant, expression can be obtained for the centre of the hypersphere which bounds the errors in the fraction.

The JVF converges when the radius of the error hypersphere $\rho_N(\mathbf{Z})$ in Eq. 26 goes to zero as N goes to infinity. For this it is necessary that **Z** lies outside the boundary of $\Gamma$ so that $\mathbf{y} \cdot \mathbf{P}_N(\mathbf{Z})$ is non-zero. For those **Z**, $\mathbf{P}_N(\mathbf{Z})$ increases exponentially with N because the zeros of $\mathbf{P}_N(\mathbf{Z})$ all lie in the boundary of $\Gamma^+$. Provided that the $\{\beta_N\}$ do not grow fast enough to cancel the growth of the

{$P_N(Z)$}, the infinite fraction converges.

### 7. Examples of Vector Fractions in Three or Two Dimensions.

We shall suppose that the shift vectors {$A_n$} of a JVF lie in a two dimensional space **X** with axes parallel to the unit vectors $x_1$ and $x_2$ and that **y** denotes a unit embedding eigenvector orthogonal to **X**. The first step in considering the convergence of such a JVF is to consider the related JVF in which the {$A_n$} are replaced by their limiting values for large n, which provide the examples discussed below.

The simplest example is a JVF with constant $A_n$ which we can make zero by shifting the origin of **Z**. If we use cylindrical polar coordinates r, y, $\phi$ it can be seen from symmetry that the asymptotic JVF does not depend on the azimuthal coordinate $\phi$ which can be fixed. Since 1/**Z** is a vector which lies in the [r, y]-plane the dimensionality has effectively been reduced to 2 and with an inverse $1/[r, y] = [r, -y]/(r^2+y^2)$ which is equivalent to the inverse of the complex variable u=r+iy. It follows that the behaviour of the asymptotic JVF in this case reduces to that of an equivalent scalar continued fraction in the complex variable u.

Next we consider shifts which have a period p,

$$A_{n+p} = A_n, \quad \beta_{n+p} = \beta_n. \tag{27}$$

We shall show that the limit of an infinite JVF can be considered as the attractive fixed point [12] of a function involving a p-level JVF.

Consider the function obtained by setting N=p in Eq. 20,

$$F(Z, T) = \beta_p R_p(Z, T), \tag{28}$$

where **Z** has a non-zero component of the embedding eigenvector. Starting with

-14-

**T=0** we now consider the successive iterations **F(Z, 0)**, **F**$^{(2)}$**(Z, 0)**,… **F**$^{(N)}$**(Z, 0)**, ... where a superscript in brackets denotes the number of times that the function is composed with itself.  It can be seen that this sequence consists of finite JVFs with p, 2p, ..., Np, ... levels.  If the infinite fraction converges to **R**$_\infty$**(Z, 0)**, it follows that,

$$\beta_p \, \mathbf{R}_\infty(\mathbf{Z, 0}) = \mathbf{F}(\mathbf{Z}, \mathbf{R}_\infty(\mathbf{Z, 0})), \tag{29}$$

so **R**$_\infty$**(Z, 0)** is a fixed point of the iterated composition **F**.  Since the successive iterations are confined to smaller and smaller spheres in **T** space it follows that this fixed point is an attractive one [12].  Equation 29 can have multiple solutions, of which only the value of the JVF is an attractive fixed point of the iterated composition.  The other solutions are repulsive fixed points and can be interpreted as continuations of the JVF, analogous to the analytic continuation of complex continued fractions to additional Riemann sheets.

When considering the convergence of JVF's in 3 dimensions, we note that the discussion in Sec. 6 applies to points which are situated off the **X**-plane.  If **Z** and **T** are in the **X**-plane however, we can represent them, together with the shifts, by complex numbers z, t and a$_n$ instead of two dimensional vectors with real components. The inverse now involves the complex conjugate, c.c.

$$1/[z_1, z_2] = [z_1, z_2]/(z_1^2 + z_2^2) = \text{c.c.}(1/(z_1 + i\, z_2)) \tag{30}$$

and what is a JVF in **Z**-space becomes an ordinary continued fraction, not of Jacobi type because it depends on both z and c.c.(z).  The simplifying feature of this is that the levels of this continued fraction are Möbius transformations in either t or c.c.(t), depending on whether p is even or odd, respectively.  The Möbius transformation f replacing **F** (or f$^{(2)}$ replacing **F**$^{(2)}$ for odd values of p) is a linear fractional transformation in t whose coefficients determine the nature of the

-15-

convergence [13]. It thus becomes a simple matter to decompose the **X**-plane into regions for which the continued fraction is oscillatory or convergent. This behaviour in the **X**-plane turns out to be related to the behaviour of the JVF when **Z** lies outside the **X**-plane.

When **Z** is outside the **X**-plane the convergence of a particular JVF can be illustrated by calculating $\rho_N(\mathbf{Z})$ defined in Eq. 26, which in turn requires the calculation of $\mathbf{P}_N(\mathbf{Z})$. The first step is to replace Eq. 17 by an equation which is linear in the polynomials. This can be done by inserting the right hand side of Eq. 16 into the right hand side of Eq. 17 and dividing the result by $p_N(\mathbf{Z})$. This yields

$$p_{N+1}(\mathbf{Z}) = |\mathbf{Z} - \mathbf{A}_N|^2 p_N(\mathbf{Z}) + \beta_N^2 p_{N-1}(\mathbf{Z}) - 2\beta_N (\mathbf{Z}-\mathbf{A}_N) \cdot \mathbf{P}_N(\mathbf{Z})^*, \qquad (31)$$

where **Z**- space is taken to have an inner product denoted by '·' in which $\mathbf{x}_1$, $\mathbf{x}_2$, and **y** are orthonormal. This equation together with Eq. 16 and the initial conditions can be used to calculate $\mathbf{P}_N(\mathbf{Z})$ over many orders of magnitude.

The calculations have been performed for one particular periodic JVF with a three-fold symmetric set of shifts defined by,

$$\mathbf{A}_0 = a\,\mathbf{x}_1; \quad \mathbf{A}_1 = a\{\cos(2\pi/3)\,\mathbf{x}_1 + \sin(2\pi/3)\,\mathbf{x}_2\};$$

$$\mathbf{A}_2 = a\,[\cos(4\pi/3)\,\mathbf{x}_1 + \sin(4\pi/3)\,\mathbf{x}_2];$$

$$\mathbf{A}_{n+3} = \mathbf{A}_n;\ a=0.4;\ \text{and}\ \beta_n=1/4. \qquad (32)$$

The upper bound for the Euclidean distance between $\mathbf{R}_N(\mathbf{Z}, \mathbf{0})$ and $\mathbf{R}_\infty(\mathbf{Z}, \mathbf{0})$ is the diameter $2\rho_N(\mathbf{Z})$ of the allowed sphere of **R**-vectors. To compare this with a numerical estimate we have performed direct calculations of Eq. 20 for various values of N with **T**=**0**. Instead of $\mathbf{R}_\infty(\mathbf{Z}, \mathbf{0})$ we have used $\mathbf{R}_M(\mathbf{Z}, \mathbf{0})$, where M is a large number chosen so that $|\mathbf{R}_n(\mathbf{Z}, \mathbf{0})-\mathbf{R}_M(\mathbf{Z}, \mathbf{0})|/|\mathbf{R}_M(\mathbf{Z}, \mathbf{0})|$ is below the floating point error for n>M. Figure 3 provides an example of the results in logarithmic form, for a point **Z** at which the convergence is sufficiently slow that the result remains larger than the rounding error for many levels. It can be seen that both the directly estimated error and the upper bound decay exponentially with N. One advantage of the latter estimate is that it can be obtained for much larger values of N because it is not limited to the rounding error in the subtraction of two numbers of similar magnitude. On the other hand it diverges as 1/**Z**·**y** near the **X**-plane. The numerical error shows no change in its qualitative behaviour near the **X**-plane, provided the JVF converges on the **X**-plane.

In the complex scalar case there are regions of the real line across which the the imaginary part of the infinite continued fraction is discontinuous. This behaviour allows the Jacobi continued fraction to be used to approximate distributions which depend on a single real variable[7]. It is therefore of interest to look for analogous behaviour in our three dimensional example.

Figures 4 and 5 show the magnitude of the **y**-component of the JVF just under the **X**-plane. The sign would be reversed at points just above the **X**-plane. This provides evidence of a discontinuity across the **X**-plane. The four dotted loops in Fig. 4 refer to the **X**-plane itself and are the boundary between convergent, outside the loops, and oscillatory behaviour, inside the loops, of the complex continued fraction obtained from Eq. 30. The fact that the infinite JVF $\mathbf{R}_\infty(\mathbf{Z}, \mathbf{0})$ is undefined at $\mathbf{Z}\cdot\mathbf{y}=0$ inside the loops is consistent with a possible discontinuity of $\mathbf{R}_\infty(\mathbf{Z}, \mathbf{0})\cdot\mathbf{y}$ when the **X**-plane is traversed.

In addition to the four main lobes, it can be seen that there is an additional isolated peak which lies outside the four dotted loops. This can be understood by reference to a different kind of behaviour on the **X**-plane. Instead of looking for oscillatory behaviour, we consider the points at which $\mathbf{R}_\infty(\mathbf{Z}, \mathbf{0}) = \infty$. Substituting $\mathbf{R}_\infty(\mathbf{Z}, \mathbf{0}) = \infty$ into Eq. 29 shows that the condition for a fixed point at infinity is equivalent to $\mathbf{R}_2(\mathbf{Z}, \mathbf{0}) = \infty$ which occurs for two values of $\mathbf{Z}$. At one of these $\mathbf{R}_\infty(\mathbf{Z}, \mathbf{0}) = \infty$ is an attractive fixed point and hence the value of the JVF at that point is $\infty$, as illustrated by the isolated spike in Fig. 4 and 5. At the other value of $\mathbf{Z}$, $\mathbf{R}_\infty(\mathbf{Z}, \mathbf{0}) = \infty$ is a repulsive fixed point and so is an example of another sheet of the continuation of $\mathbf{R}_\infty(\mathbf{Z}, \mathbf{0})$.

**8. Quadrature and Geometry**

A strong motivation for attempts to generalise the properties of ordinary continued fractions and orthogonal polynomials is the search for multi-dimensional quadrature formulas, generalisations of Gaussian quadrature. We have considered quadratures in relation to the work described above, and offer some comments below.

Finite JVFs do have discrete, oriented zeros and infinities like the quadrature points generated by finite Jacobi continued fractions. However, we have constructed examples of JVFs for which the numbers of zeros and infinities do not each increase by one with each additional level of the fraction. In this regard, JVFs differ from the Jacobi continued fractions associated with Gaussian quadrature.

We have noted a difference between one-dimensional and multi-dimensional quadratures, which may be important. One construction of a Gaussian quadrature is to minimise the integral of a positive weight distribution over a quadratic form which is the squared magnitude of a polynomial [14]. The zeros of this minimal polynomial are then the quadrature points, and the weights are given by integrals over polynomials of lower degree. For one-dimensional integrals, the minimal polynomial always has the maximum number of distinct zeros; while for multi-dimensional integrals, minimal multi-nomials do not.

Whether this is related to the above deficiencies in the numbers of zeros and infinities of JVFs is not clear.

There is current interest in the construction of geometric algebras, for example Clifford Algebras [15]. Our last comment is to point out that the introduction of an inverse vector used in this work is a very simple example of a geometric algebra.

**Acknowledgements**

The hospitality of Pembroke College and the Cavendish Laboratory in Cambridge together with that of Gelly Farm, West Glamorgan, and financial support from the Richmond F. Snyder Fund as well as the UK Engineering and Physical Sciences Research Council (GR/S03263/01) are gratefully acknowledged by RH. GW is grateful to the Faculty of Technology of the Open University, Milton Keynes, for support during the early stages of this work.

**References**

1. L. Lorentzen and H. Waadeland, *Continued Fractions with Applications*, (North Holland, Amsterdam, 1992), also W.B. Jones and W.J. Thron, *Continued Fractions: Analytic Theory and Applications*, Encyclopedia of Mathematics and its Applicaitons **11** (Addison-Wesley, Reading, Mass. 1980).

2. T.S. Chihara, *An Introduction to Orthogonal Polynomials*, (Gordon and Breach, New York, 1978).

3. J.A. Shohat and J.D. Tamarkin, *The Classical Moment Problem*, Math. Surv. I, rev. ed. (Am. Math. Soc., Providence, Rhode Island, 1950), pp. 106-26

4. A. Ben-Israel and T.N.E. Grenville, *Generalized Inverses: Theory and Applications*, (Wiley, New York, 1980).

5. F. Klein, *Elementary Mathematics from an Advanced Standpoint*, 3rd. Ed. (Macmillan, New York,1939) p.100.

6. J.L.Coolidge, *A Treatise on Algebraic Curves*, ( Clarendon Press, Oxford, 1931).

7. Reference 3, pp. 23-7.

8. *Encyclopaedia of Mathematics,* vol. 2, ed. M. Hazewinkle, (Kluwer, Dortdrech, 1988), p316, 320.

9. N.I. Akhiezer, *The Classical Moment Problem*, (Oliver and Boyd, Edinburgh, 1965), page 8.

10. Reference 2, p.23.

11. Reference 3, p.48.

12. R.L. Devaney, *An Introduction to Chaotic Dynamical Systems*, (Benjamin/Cummings, New York, 1986) p. 27.

13. G. Sansone and J. Gerritsen, *Lectures on the Theory of Functions of a Complex Variable*, ( Wolters-Noordhoff, Groningen, 1969) Vol.2, p.24.

14. Reference 2, page 17, Exercise 3.4.

15. D. Hestenes and G. Sobczyk, *Clifford Algebra to Geometric Calculus*, (D. Riedel, Dortrecht, 1984) and *Clifford (Geometric) Algebras*, ed. W.E. Baylis (Birkhäuser, Boston, 1996).

Figure Captions

1. Schematic illustration of the mapping from **T**-space (shaded) to nested, kissing, hyperspheres of $\mathbf{R_N}(\mathbf{Z}, \mathbf{T})$-space (white) describing the convergence of a JVF.
2. How the maximum and minimum of $|\mathbf{R_N}(\mathbf{Z}, \mathbf{T})|$ determine the diameter of the hypersphere in $\mathbf{R_N}(\mathbf{Z}, \mathbf{T})$-space (white) which is the image of $\Gamma$ from **T**-space (shaded).
3. Comparison of $\log_{10}(2\rho_N)$ (dotted) with $\log_{10}|\mathbf{R}_N - \mathbf{R}_{2000}|$ (crosses) vs number of levels N of periodic JVF evaluated at the point
$\mathbf{Z} = (-0.26)\ \mathbf{x}_1 + (0.69)\ \mathbf{x}_2 + (0.001)\ \mathbf{y}$.
4. Contour diagram for $\mathbf{R}_{3000} \cdot \mathbf{y}(x_1, x_2, -0.001)$. The contours should be counted inwards, starting from n=0 at the outermost solid line on each of the five lobes. The heights of the contours are then specified to be $y_n = y_0\ e^{k\,n}$ where n = 0,1,2... $y_0$=.0864 and k=0.312544. It is expected that the contours will not change much in the limit N→∞ as this figure cannot be distinguished from a similar one with 6000 levels. If this limit were to be followed by a second one namely y→0, we should expect that the tallest peak would sharpen to a point singularity while the other sets of contours would be confined to within the four dotted loops.
5. $\mathbf{y} \cdot \mathbf{R}_{3000}(x_1, x_2, -0.001)$ versus $x_1$ and $x_2$. The figure is an orthographic projection of the contour map shown in Fig. 4. The peak with the flat top has been truncated by reducing the vertical range; its true value is 142.37.

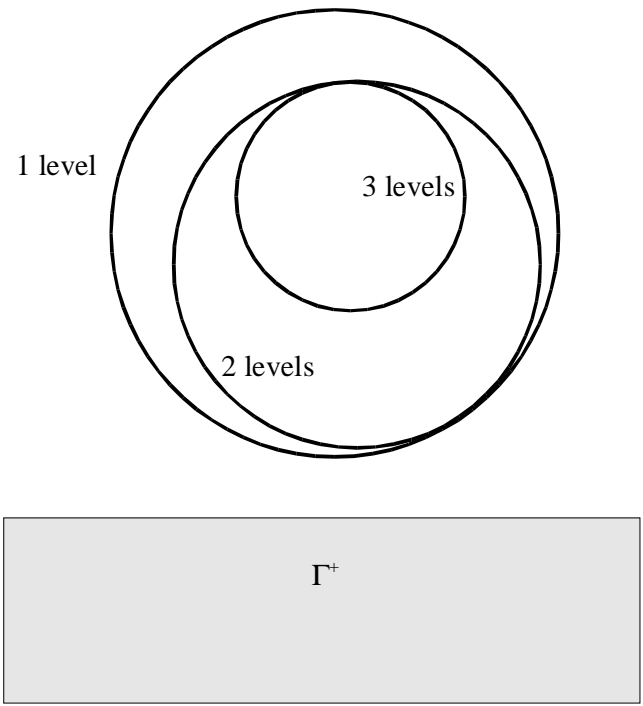

Figure 1

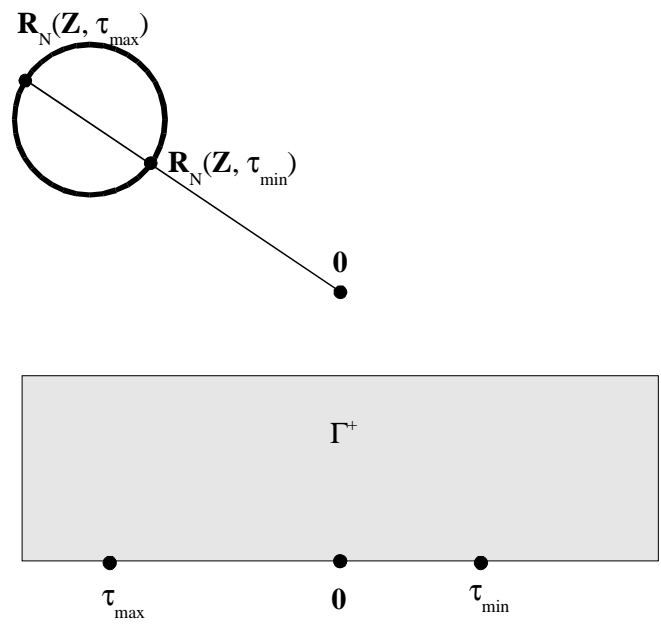

Figure 2

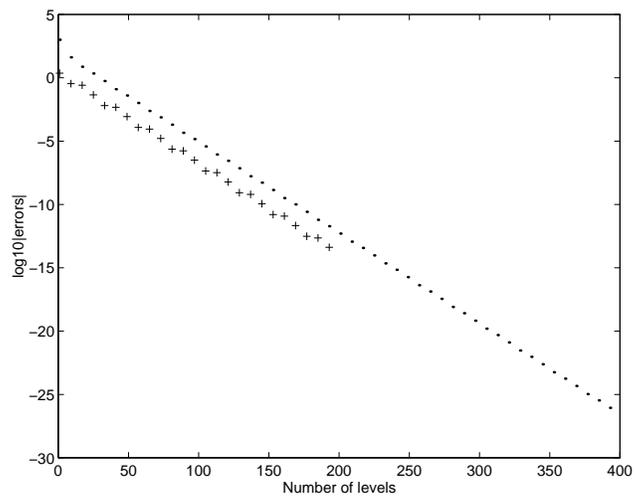

Figure 3.

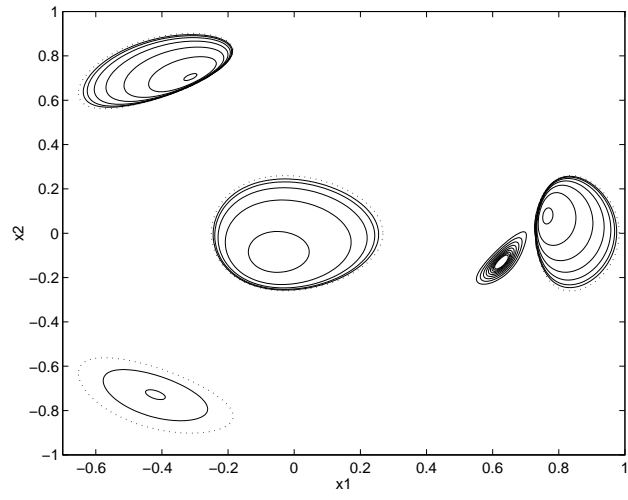

Figure 4.

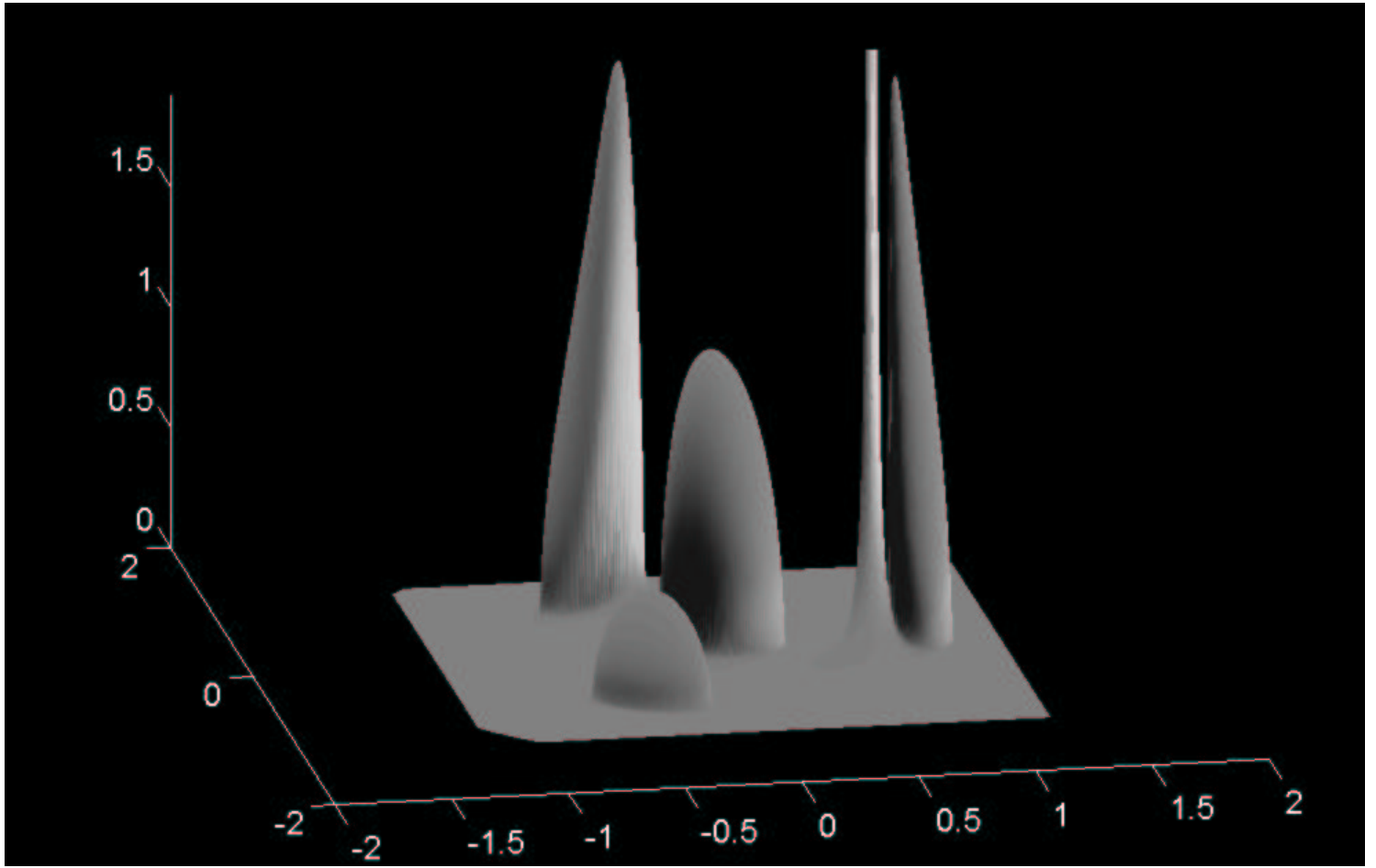

Figure 5.